\documentclass[onecolumn]{aastex63}
\usepackage{amsmath}
\usepackage{amssymb}
\usepackage{graphicx}
\usepackage{float}
\usepackage[T1]{fontenc}
\usepackage{multirow}
\usepackage{appendix}
\usepackage{xcolor}
\colorlet{purple}{blue!40!red}
\colorlet{orange}{red!40!yellow}
\usepackage{color}
\definecolor{darkgreen}{RGB}{0,120,0}

\newcommand{\rmC}{\mathrm{C}}
\newcommand{\rmS}{\mathrm{S}}
\newcommand{\rmL}{\mathrm{L}}

\begin{document}

\title{Modeling Galaxy Surveys with Hybrid SBI}

\correspondingauthor{Gemma Zhang}
\author{Gemma Zhang}
\email{yzhang7@g.harvard.edu}
\affiliation{Department of Physics, Harvard University, Cambridge, MA 02138, USA}
\affiliation{The NSF AI Institute for Artificial Intelligence and Fundamental Interactions}

\author{Chirag Modi}
\affiliation{Center for Cosmology and Particle Physics, New York University, New York, NY 10003, USA}

\author{Oliver~H.\,E.~Philcox}
\affiliation{Simons Society of Fellows, Simons Foundation, New York, NY 10010, USA}
\affiliation{Center for Theoretical Physics, Columbia University, New York, NY 10027, USA}
\affiliation{Department of Physics,
Stanford University, Stanford, CA 94305, USA}

\begin{abstract}
    \noindent Simulation-based inference (SBI) has emerged as a powerful tool for extracting cosmological information from galaxy surveys deep into the non-linear regime. Despite its great promise, its application is limited by the computational cost of running simulations that can describe the increasingly-large cosmological datasets. Recent work proposed a hybrid SBI framework (HySBI), which combines SBI on small-scales with perturbation theory (PT) on large-scales, allowing information to be extracted from high-resolution observations without large-volume simulations. In this work, we lay out the HySBI framework for galaxy clustering, a key step towards its application to next-generation datasets. We study the choice of priors on the  parameters for modeling galaxies in PT analysis and in simulation-based analyses, as well as investigate their cosmology dependence. By jointly modeling large- and small-scale statistics and their associated nuisance parameters, we show that HySBI can obtain 20\% and 60\% tighter constraints on $\Omega_m$ and $\sigma_8$, respectively, compared to traditional PT analyses, thus demonstrating the efficacy of this approach to maximally extract information from upcoming spectroscopic datasets.
\end{abstract}

\keywords{Cosmology (343), Large-scale structure of the universe (902), Astrostatistics (1882)}

\section{Introduction}
\noindent 
The distribution of galaxies measured using large astronomical surveys has offered us deep insights into the composition and evolutionary history of our Universe. Ongoing and upcoming experiments, such as the Dark Energy Spectroscopic Instrument (DESI)~\citep{desi2016part1, desi2016part2,desi2022overview}, the \textit{Euclid} satellite mission~\citep{euclid2011}, the SPHEREx observatory~\citep{SPHEREx:2014bgr}, the Nancy Grace Roman Space Telescope~\citep{wfirst2015, roman2022}, and the Rubin Observatory Legacy Survey of Space and Time (LSST)~\citep{lsst2019}, will provide us with data of unprecedented volume and resolution, covering increasingly large volumes of our Universe. Such a wealth of information will lead to significant improvements in our understanding of the standard model; however, it remains unclear how to fully utilize this data to obtain the tightest constraints on cosmological parameters.

Recent analyses of galaxy clustering have relied primarily on perturbation theory (PT), which constrains cosmological parameters through analytical computation of low-order summary statistics, including the two- and three-point correlation functions~\citep[e.g.,][]{philcox2022boss, damico2024boss, chen2022new, DESI:2024jis,Chen:2024vuf}. While these methods provide accurate models of the galaxy distribution in the linear and quasi-linear regime, they naturally break down on smaller scales due to the inherent non-perturbativity of the density and velocity fields. Moreover, computing the theoretical models for non-Gaussian statistics beyond leading-order can itself be highly non-trivial \citep{Philcox:2022frc,Anastasiou:2022udy}. Recently, a number of works have explored supplementing perturbative approaches with numerical simulations, in particular, using them to place a prior on the nuisance parameters entering the PT description~\citep{ivanov2024fullshape, ivanov2024fullshape2, zhang2024hodinformed,Ivanov:2024dgv}. This could yield significantly sharper constraints, though one is still restricted to the large-scale regime. To fully exploit the information content of galaxy surveys, we require alternative methods that are capable of extracting constraints from small scales where highly non-linear structure formation comes into play \citep[e.g.,][]{Beyond-2pt:2024mqz}. 

With the advent of machine learning and high-resolution N-body simulations, simulation-based inference (SBI) has emerged as a promising alternative to PT \citep{papamakarios2016fast, alsing2018massive, cranmer2020frontier, Jeffrey2020, hahn2023simbig, Modi23_sensitivity}, capable of leveraging information from galaxy clustering in the non-linear regime. In general, SBI starts out with a large set of simulated observations, each produced with a given set of input parameters. It then uses neural density estimation to learn the relationships between the observables and their underlying parameters from these simulations. To use SBI in cosmological analyses, we require a set of simulations that (a) match real observations in size, (b) provide sufficient coverage of the parameter space of interest, and (c) are appropriately accurate. 

With current and upcoming surveys probing increasingly large spatial volumes (and increasingly small scales), producing suites of high-resolution simulations that can be used for SBI will be difficult. To address this, \cite{modi2023hybridsbi} introduced Hybrid SBI (HySBI), a hybrid approach that bridges PT and SBI to leverage information from a wide range of scales while circumventing the need to run full-size simulations. This was achieved by combining the PT-based statistics and analytic likelihood on large scales with conditional SBI inference on small scales, thus requiring only small `subbox' simulations. Using summary statistics computed from the dark matter density fields of N-body simulations, they demonstrated that HySBI can lead to significant computational gains with minimal loss of inference performance. 

Though encouraging, the aforementioned methods cannot yet be applied to data, since three-dimensional dark matter density fields are not direct observables from cosmological surveys. In this paper, we extend the work of \cite{modi2023hybridsbi} by applying HySBI to summary statistics of galaxy clustering, emulating the observables measured by, for example, DESI. This is non-trivial since, compared to the dark matter field, the galaxy distribution contains additional nuisance parameters in both large-scale and small-scale models (such as bias and halo occupation distribution parameters). Furthermore, the parameters governing these two scales have inter-dependence that should be taken into account if we seek to maximize the cosmological information we can extract from this data. In this work, we address these complexities and develop a hybrid analysis pipeline that can be applied to redshift-space galaxy clustering data. This is a key step towards using HySBI to perform cosmological inference using observations from next-generation telescopes. 

\vskip 20pt
The remainder of this work is organized as follows. In Sec.~\ref{sec:method}, we introduce our HySBI method in the context of the redshift-space galaxy power spectrum. In Sec.~\ref{sec:simulation}, we provide details about the simulations that are used for training SBI and as mock observations for inference. We present and discuss our main results in Sec.~\ref{sec:results} and conclude in Sec.~\ref{sec:conclusion}. Additional results are shown in Appendix \ref{app}.


\vskip 20pt
\section{Methodology: Hybrid Simulation-based Inference}
\label{sec:method}
\noindent Given cosmological observations (data, $x$), our goal is to infer the cosmological parameters $\theta_{\mathrm{C}}$ (e.g., $\Omega_m,H_0,\sigma_8,\cdots$), \textit{i.e.}~we are interested in the posterior distribution $p(\theta_{\mathrm{C}} | x)$. In many cases, we can decompose the data into a large-scale component $x_{\mathrm{L}}$ that lies in the quasi-linear regime and can be modeled analytically, and small-scale observations $x_{\mathrm{S}}$ that are non-linear and hence can only be modeled using simulations. 
Under this decomposition, the posterior distribution can be written as follows, using Bayes' Theorem:
\begin{align}
    p(\theta | x_{\mathrm{S}}, x_{\mathrm{L}}) \propto p_{\mathrm{PT}}(x_{\mathrm{L}}|\theta_{\mathrm{C}}, \theta_{\mathrm{L}}) p_{\mathrm{sim}}(x_{\mathrm{S}}| \theta_{\mathrm{C}}, \theta_{\mathrm{S}}, x_{\mathrm{L}}) p(\theta)
    \label{eq:bayes}.
\end{align} 
Here, the total parameter set $\theta := (\theta_{\mathrm{C}},\, \theta_{\mathrm{L}},\, \theta_{\mathrm{L}})$ encompasses cosmological parameters ($\theta_{\rm C}$), large-scale PT-based modeling parameters ($\theta_{\rm L}$), and small-scale simulation modeling parameters ($\theta_{\rm S}$).
Note that this decomposition applies generically to any observable and model choice, provided we can make some phenomenological split into large and small scales. For instance, $x_{\rm L}$ can be any large-scale statistic such as power spectra or bispectra, and $x_{\mathrm{S}}$ can be small scale correlation functions or more sophisticated summary statistics like wavelet statistics, Minkowski functionals or even learned-neural summary statistics.  

In this work, we will focus on galaxy clustering data. 
In this setting, $\theta_{\mathrm{L}}$ corresponds to the set of galaxy-bias parameters from the Effective Field Theory of Large Scale Structure (hereafter EFT) while $\theta_{\mathrm{S}}$ represents halo-occupation distribution (HOD) parameters. For the large- and small-scale statistic, we use the large- and small-scale power spectra. We discuss how to model these in Sec.~\ref{sec:PT} using EFT and Sec.~\ref{sec:SBI} using SBI respectively. 
An important consideration is the choice of prior $p(\theta) = p(\theta_{\mathrm{C}}, \theta_{\mathrm{L}}, \theta_{\mathrm{S}})$; this is discussed in Sec.~\ref{sec:hybrid_sbi}.

\subsection{Large-Scales: Perturbation Theory}
\label{sec:PT}
\noindent The EFT of large-scale structure serves as our theory model for the large-scale galaxy power spectrum. This is computed using the Boltzmann code extension \texttt{CLASS-PT}~\citep{chudaykin2020nonlinear}, which outputs a set of 48 theoretical power spectrum components, $g_n(\theta_{\rm C})$, which we then combine with the set of nuisance parameters $\theta_{\rm L}$:
\begin{align}
    x_{\rm theory} = \sum_{n=1}^{48}f_n(\theta_{\rm L})g_n(\theta_{\rm C}).
\end{align}
The precise expressions for $g_n$ and $f_n$ 
can be found in \citet{chudaykin2020nonlinear,philcox2022boss}. 
Explicitly, $\theta_{\rm L}$ is given by
\begin{align}
    \{b_1,b_2,b_{\mathcal{G}_2},b_{\Gamma_3}\}\times \{c_0,c_2,b_4\}\times\{P_{\rm shot},a_0,a_2\}
\end{align}
where the first set encodes biases, the second counterterms (from integrating-out small-scale physics, such as the Fingers-of-God effect), and the third stochasticity, such as shot-noise. We choose the following conservative PT priors on these EFT parameters: 
\begin{gather*}\label{eq: pt-prior}
    b_1 \sim \mathcal{U}(0, 4) \quad b_2 \sim \mathcal{N}(0, 3) \quad b_{\mathcal{G}_2} \sim \mathcal{N}(0, 3) \quad b_{\Gamma_3} \sim \mathcal{N}(-1, 4) \\
    c_0 \sim \mathcal{N} \left( 0, 150 \right) \quad c_2 \sim \mathcal{N} \left( 30, 150 \right) \quad b_4 \sim \mathcal{N} \left( 500, 500 \right)\\ 
    P_{\text{shot}} \sim \mathcal{N} \left( \frac{\bar{n}_0}{\bar{n}}, \frac{\bar{n}_0}{\bar{n}} \right) \quad a_0 \sim \mathcal{N} \left( 0, \frac{2.5 \bar{n}_0}{\bar{n}} \right) \quad a_2 \sim \mathcal{N} \left( 0, \frac{2 \bar{n}_0}{\bar{n}} \right), 
\end{gather*} 
where we scale the shot-noise with respect to the specific number density of the sample, $\bar{n}$, normalizing with respect to $\bar{n}_0 = 5 \times 10^{-4}$~Mpc$^{-3}$ based on the luminous red galaxy density projected for the full DESI sample~\citep{zhou2023target}. These priors are similar to those used in analyses of BOSS clustering data \citep{philcox2022boss}, but are slightly enlarged to ensure that we encompass the wide range of bias parameter values measured from simulations 
(see Sec.~\ref{sec:hybrid_sbi}). 

To expedite computation, we first run \texttt{CLASS-PT} using a wide set of cosmologies before interpolating the output power spectrum components using a set of emulators. These are based on a simple multilayer perception architecture and are trained with a mean squared error (MSE) loss function. As in previous works (and validated in~\citet{Tucci:2023bag}), we use a Gaussian likelihood to model the difference in power spectra between observations $x_{\rm L}$ and our theory forward model $x_{\mathrm{theory}}$: 
\begin{align}
    \label{eq:pt_like}
    p_\mathrm{PT}(x_{\rm L} | \theta_\rmC, \theta_\rmL) \propto \exp{\left(-\frac{1}{2}(x_{\rm L} - x_{\mathrm{theory}})^T C^{-1} (x_{\rm L} - x_{\mathrm{theory}})\right)} 
\end{align}
where $C$ is the covariance matrix of the binned power spectrum. To account for redshift-space distortions (RSD), we include both the monopole and quadrupole moments of the galaxy power spectra, and assume an idealized Gaussian covariance, which is diagonal in $k$ \citep[following][]{chudaykin2019measuring}. 

\subsection{Small-Scales: Simulation-based Inference}
\label{sec:SBI}
\noindent Since the PT models fail on small scales due to increased non-linearity, we use simulation-based inference to extract information from this regime. 
A typical SBI involves two steps--- (1) generating a training dataset of $(\theta, x)$ for model parameters to be inferred ($\theta$) and corresponding data ($x$) over the prior range ($p(\theta)$), and then (2) using flexible neural density estimators to learn the likelihood distribution $p(x|\theta)$ or the posterior distribution $p(\theta|x)$.  
Here, we are interested in learning the conditional small-scale data likelihood in Eq.~\ref{eq:bayes}--- $p(x_{\mathrm{S}}| \theta_{\mathrm{C}}, \theta_{\mathrm{S}}, x_{\mathrm{L}})$.
As such, the data corresponds to the small-scale power spectrum multipoles $x := x_\rmS$ and the parameters correspond to the cosmology parameters  and HOD parameters ($\theta :=(\theta_\rmC, \theta_\rmS)$).

Notably, the likelihood we wish to learn is conditioned on the observed large-scale data $x_\rmL$. While not present in typical SBI, this is necessary to correctly account for the covariance between the two regimes and requires us to modify the training of SBI. 
HySBI seeks to learn $p(x_{\mathrm{S}}| \theta_{\mathrm{C}}, \theta_{\mathrm{S}}, x_{\mathrm{L}})$ using customized small-volume simulations without running the survey size simulations at full fidelity as required by traditional SBI.
\citet{modi2023hybridsbi} point out that the training data for learning this conditional distribution can be generated by running computationally inexpensive, approximate simulations in full-volume to simulate $x_\rmL$, and then simulating only a sub-volume with high-fidelity simulations to model ($x_\rmS|x_\rmL$).
In the absence of such hybrid simulations existing in the community, we follow~\cite{modi2023hybridsbi} to emulate this setup by cutting a large simulation box into subboxes and measuring $x_\rmS$ on these subboxes only while $x_\rmL$ is measured in the full-box. To account for boundary conditions when computing the subbox power spectra, we correct the measured power spectra $\tilde{P}_\ell(k)$ with a window function matrix $M$ such that $P_\ell(k) = [M^{-1}\tilde{P}]_\ell(k)$, where $P_\ell(k)$ is the true power spectra and $\ell\in\{0,2\}$ are the monopole and quadrupole moments.
We estimate $M$ numerically by injecting white-noise into each $k,\ell$-bin in turn and measuring its leakage to other bins. To minimize stochastic effects, we average $M$ over 50 realizations (using \texttt{nbodykit}).
Finally, we average of $x_\rmS$ over multiple subboxes to minimize the effect of super-sample variance. We provide details in Sec.~\ref{sec:nn} regarding the training of neural density estimators using these power spectra as input data. 


\subsection{Combined Scales: Joint Priors}
\label{sec:hybrid_sbi}

\noindent 
The final ingredient to obtain the posterior in Eq.~\ref{eq:bayes} is the prior distribution $p(\theta) = p(\theta_{\mathrm{L}}, \theta_{\mathrm{C}}, \theta_{\mathrm{S}})$ where, for galaxy clustering, $\theta_\rmL$ and $\theta_\rmS$ are the bias parameters and HOD parameters. 
Typically one assumes broad, independent priors on these parameters. Here, we will assume a uniform prior on both the cosmological and small-scale parameters $\theta_{\mathrm{C}}$ and $\theta_{\mathrm{S}}$ within the parameter upper and lower bounds from the training set. For $\theta_\rmL$, we explore two different options. First, we assume that the small- and large-scale parameters are independent, adopting the conservative EFT prior on $p(\theta_{\rm L})$ given in Eq.~\ref{eq: pt-prior} 
\citep[e.g.,][]{philcox2022boss}.
However, this is sub-optimal since both bias and HOD parameters describe the same underlying galaxy distribution and are thus related for a given cosmology. To maximize the information content of our hybrid analysis, we can decompose the prior distribution: $p(\theta_{\mathrm{L}}, \theta_{\mathrm{C}}, \theta_{\mathrm{S}}) = p(\theta_{\mathrm{L}} | \theta_{\mathrm{C}}, \theta_{\mathrm{S}}) p(\theta_{\mathrm{C}}, \theta_{\mathrm{S}})$. Here, the conditional prior $p(\theta_{\mathrm{L}} | \theta_{\mathrm{C}}, \theta_{\mathrm{S}})$ describes the correlation between large- and small-scale parameters and can be computed using simulations.

To learn the joint prior, we generate galaxy catalogs for simulations with different cosmologies using an HOD model, as described in the next section. For each simulated catalog, we fit for the bias parameters $b_1, b_2, b_{\mathcal{G}_2}, b_{\Gamma_3}$ directly from the field-level density field using the methodology developed in~\cite{ivanov2024fullshape,ivanov2024fullshape2}. This utilizes the known initial conditions of the simulations to obtain tight constraints on biases through a field-level likelihood obtained using Zel'dovich shifted operators, as discussed in \citet{Schmittfull:2018yuk}. Given the bias parameters for each realization, we fit for the counterterms and stochastic terms by minimizing the residual in galaxy power spectra, noting that these are harder to estimate from the field-level simulations (though see \citealt{ivanov2024fullshape2}).
With this dataset of cosmology parameters, HOD parameters and corresponding bias parameters, we train normalizing flows to learn the conditional prior distribution.

Finally, in addition to the independent priors and simulation-informed priors, we also experiment with hybrid priors that combine these two approaches. These are discussed in Sec.~\ref{sec:results} in more detail. 





\section{Simulations, Summary Statistics and Training}
\label{sec:simulation}

\noindent In this section, we describe the simulated data used in this work, how the statistics are estimated, and how different normalizing flow architectures are trained for SBI and learning the prior distribution. 

\subsection{Simulations}
\noindent In this work, we use the \textsc{Quijote}~\citep{Villaescusa-Navarro2020quijote} high-resolution $\Lambda$CDM Latin-hypercube (LHC) simulations, holding out a small subset as our mock observations for testing. This suite consists of 2,000 N-body simulations, each of which contains $1024^3$ cold dark matter particles in a $1(h^{-1}\mathrm{Gpc})^3$ box and corresponds to a different set of cosmological parameters: $\{\Omega_m, \Omega_b, h, n_s, \sigma_8\}$. 
We use catalogs of dark matter halos identified by the friends-of-friends algorithm. 

We populate these halos with galaxies based on an HOD model using the  \texttt{nbodykit} package~\citep{hand2018nbodykit}. Following \cite{zheng2007hod}, our HOD models the probability of a galaxy occupying a central or satellite halo by the following: 
\begin{align*}
    \langle N_{\text{cen}}(M) \rangle &= \frac{1}{2} \left[ 1 + \operatorname{erf} \left( \frac{\log M - \log M_{\text{min}}}{\sigma_{\log M}} \right) \right] \\
    \langle N_{\text{sat}}(M) \rangle &= \frac{1}{2} \left[ 1 + \operatorname{erf} \left( \frac{\log M - \log M_{\text{min}}}{\sigma_{\log M}} \right) \right] \left( \frac{M - M_0}{M_1'} \right)^{\alpha},
\end{align*}
where $\langle N_{\text{cen/sat}}(M) \rangle$ is the mean occupation function of central/satellite galaxies and $\{M_{\rm min},\sigma_{\log M},M_0,M_1',\alpha\}$ are free parameters that depend on the galaxy sample of interest. We follow \cite{hahn2023simbig} using a decorated HOD model based on~\cite{hearin2016introducing}, including two assembly bias parameters which introduce a dependence of halo occupation probability on the halo concentration. 
 We use the following ranges for our HOD model parameters following~\cite{hahn2023simbig}: 
\begin{gather*}
    \log M_{\mathrm{min}} \sim \mathcal{U}(12, 14) \quad \sigma_{\log M} \sim \mathcal{U}(0.1, 0.6) \quad \log M_0 \sim \mathcal{U}(13, 15) \quad \log M_1 \sim \mathcal{U}(13, 15) \quad \alpha \sim \mathcal{U}(0, 1.5) \\
    A_{\mathrm{bias, central}} \sim \mathcal{N}(0, 0.2)~\mathrm{over}~[-1, 1] \quad A_{\mathrm{bias, satellite}} \sim \mathcal{N}(0, 0.2)~\mathrm{over}~[-1, 1]. 
\end{gather*}
Note that this set of parameters governs the small-scale modeling in this work (\textit{i.e.}~$\theta_{\rm S} = \theta_{\rm HOD}$). 
For each cosmology in \textsc{Quijote} LHC, we generate 50 different HOD realizations and keep simulations with galaxy number densities between $2.5\times 10^{-4}$ and $10^{-3}$~Mpc$^{-3}$.
This number density range is chosen to be consistent with the luminous red galaxy density of $5 \times 10^{-4}$~Mpc$^{-3}$ in the DESI~\citep{zhou2023target}.
We also take 30 of these HOD realizations for each of the \textsc{Quijote} LHC simulations to produce the dataset for training the flow prior $p(\theta_{\mathrm{L}} | \theta_{\mathrm{C}}, \theta_{\mathrm{S}})$.

\subsection{Summary Statistics}
\noindent We use the resulting galaxy catalog to compute the mock galaxy power spectra using fast Fourier transforms. 
For the baseline PT-only analysis in this paper, we use scales $k \in (0.007, 0.2)~h\mathrm{Mpc}^{-1}$, with the upper-bound chosen to be consistent with previous works \citep[e.g.][]{philcox2022boss}.
For the hybrid analysis, we split the power spectra into large-scale and small-scale data, $x_{\rm L}$ and $x_{\rm S}$. We split at $k = 0.15~h\mathrm{Mpc}^{-1}$, so that modes with $k \in (0.007, 0.15)~h\mathrm{Mpc}^{-1}$ comprise the large-scale statistic modeled with PT and those with $k \in (0.15, 0.5)~h\mathrm{Mpc}^{-1}$ make up the small-scale statistics modeled with SBI. As noted in Sec.~\ref{sec:SBI}, for training SBI on small scales, we split our large Quijote simulation box into eight subboxes and estimate the small scale power spectra from the (aperiodic) subboxes. These are then corrected using the window function matrix $M$, and averaged together. However, we emphasize that the small scale statistics used for the test data are measured on the full simulation box, without chopping into subboxes, in order to emulate real data analysis. Finally, the width of $k$ bins is equal to the fundamental frequency of the large box for the large scale statistic, and similarly the fundamental frequency of the subbox for the small scale statistic. 


\subsection{Neural network training and inference}
\label{sec:nn}
\noindent We use normalizing flows, in particular masked autoregressive flows (MAF)~\citep{papamakarios2017maf}, to learn both the small-scale likelihood term $p(x_{\mathrm{S}}| \theta_{\mathrm{C}}, \theta_{\mathrm{S}}, x_{\mathrm{L}})$ and the simulation-informed conditional prior $p(\theta_{\mathrm{L}} | \theta_{\mathrm{C}}, \theta_{\mathrm{S}})$. The training objective is to minimize the negative log-probability over the training dataset. We implement this with the NLE functionality for the likelihood term and NPE functionality for the prior term using the \texttt{sbi} package~\citep{tejero-cantero2020sbi}. For robustness, we train 40 networks and use an ensemble of 10 networks with the best validation loss to model the desired likelihood or conditional prior. After training our normalizing flows, we sample the posterior using the \texttt{emcee} package~\citep{foremanmackey2013emcee}, varying all relevant nuisance parameters as well as the five cosmology parameters.  

To determine the architectures of our normalizing flows, we perform a hyperparameter search using the Weights \& Biases code~\citep{wandb}. We randomly vary the batch size, the number of layers in the neural network, the number of hidden features, and the learning rate in each run. To aid the training process, we standardize the power spectrum input in each $k$ bin with the standard scaler and standardize the HOD parameters with the `minmax' scaler, both from the \texttt{scikit-learn}~\citep{sklearn} package. To determine the hyperparameters for the emulators used in PT analysis, we perform a coarse search in the parameter space and chose a set of fixed values for our models. We have verified that this choice does not affect our results significantly. The code and details of our training are publicly available.\footnote{\href{https://github.com/gemyxzhang/HySBI-galaxy}{https://github.com/gemyxzhang/HySBI-galaxy}} 

\vskip 20pt
\section{Results}\label{sec:results}

\noindent Armed with our setup, we now set out to investigate the utility of HySBI in the context of future galaxy clustering surveys. Specifically, we seek to answer the following questions:
\begin{enumerate}
    \item How does the simulation-informed conditional prior on the large scale model parameters $p(\theta_{\rm L}|\theta_{\rm C},\theta_{\rm S})$ compare with commonly used conservative EFT priors? 
    \item How much more information do we gain by combining both large and small scales (HySBI) compared to analyzing only large (PT) or small (SBI) scales?
    \item How does capturing the interplay between large- and small-scale model parameters, \textit{i.e.}~using the informed prior $p(\theta_{\rm L}|\theta_{\rm C},\theta_{\rm S})$, help HySBI over using independent priors on all model parameters?
\end{enumerate}
As an aside, we also investigate if we can enhance the information content of large-scale PT only analyses by adding insights on galaxy bias parameters from small scales \citep[as in][]{ivanov2024fullshape}.

In this section, we provide answers to these questions and make recommendations for future hybrid studies. For clarity, all plots show results for a single HOD realization and cosmology (close to the center of the training set); we note that our conclusions also hold for other cosmologies examined. 

\subsection{Simulation-informed prior distribution}

\begin{figure}[h!]
    \centering
    \includegraphics[width=0.99\linewidth]{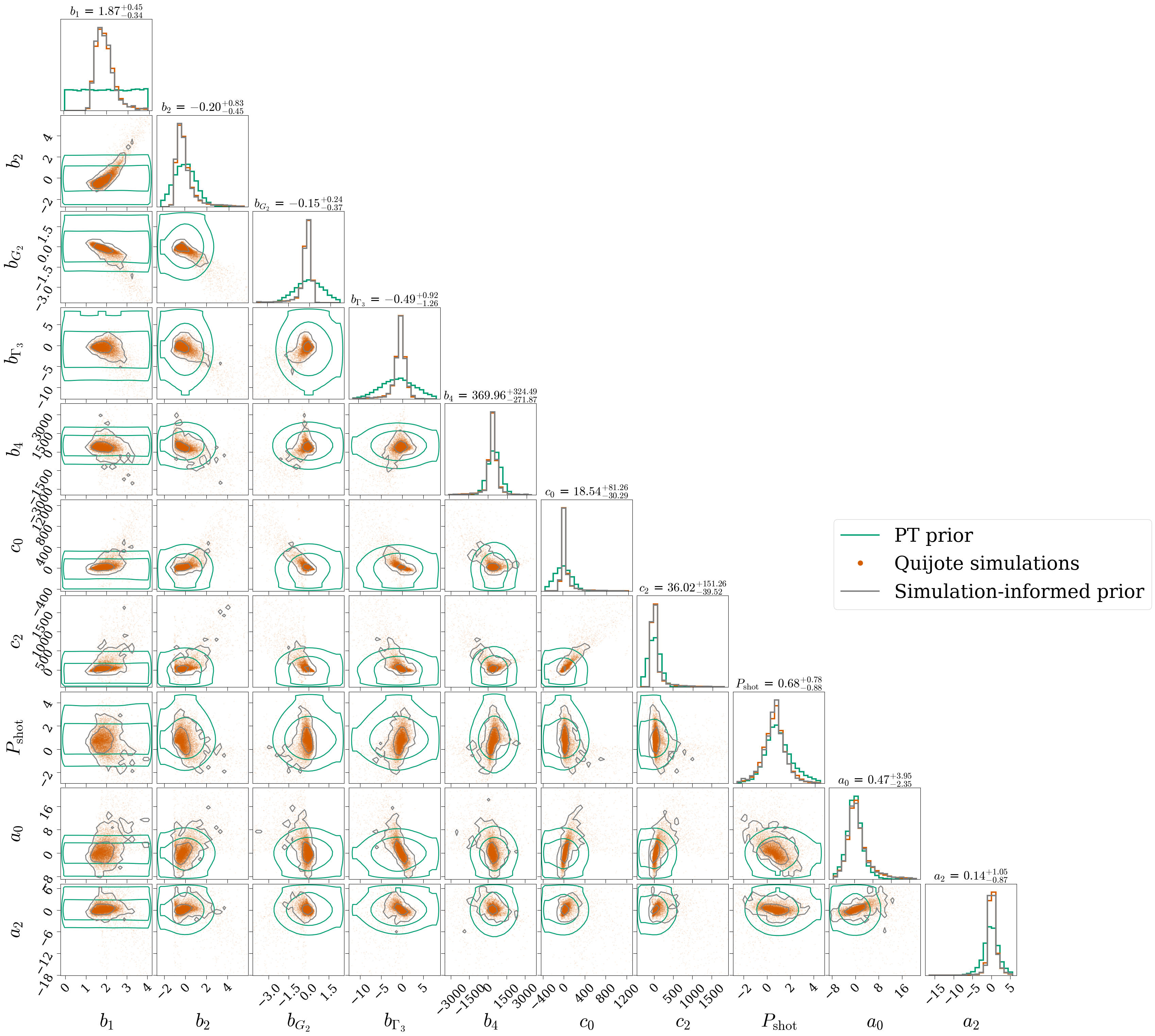}
    \caption{Bias parameters (orange) measured from HOD realizations of the 2,000 \textsc{Quijote} high-resolution LHC simulations covering a broad range of HOD and cosmology parameters. 
    The 16th and 84th percentiles of each parameter are shown above each column. Green contours show the conservative PT priors discussed in Sec.~\ref{sec:PT} and gray contours show the simulation-based prior. Despite the marginalization over cosmology and HOD parameters, we find non-trivial relationships between bias parameters, which can be used to sharpen our HySBI analyses.
    }
    \label{fig:bias_fits}
\end{figure}

\noindent In Fig.~\ref{fig:bias_fits}, we plot the distribution of the large-scale nuisance parameters (encompassing biases, counterterms, and stochasticity coefficients) learned from the simulations alongside the conservative priors on $\theta_{\rm L}$ discussed in Sec.~\ref{sec:PT}.
This is achieved by sampling from our trained simulation-informed priors while varying both cosmology and HOD parameters, thus marginalizing these out. We also plot the training set of the simulation-informed prior (\textit{i.e.} EFT parameters from simulations) in scatter points. As desired, the conservative priors in Sec.~\ref{sec:PT} have a broader coverage in parameter space than the distribution of the EFT parameters from simulations. We also find non-trivial dependence between different bias parameters and counterterms. This has been shown in previous works such as \cite{ivanov2024fullshape} and \cite{zhang2024hodinformed}. Our correlations are somewhat weaker than those of previous works due to the wide range of cosmological parameters adopted herein. 
Since the nuisance parameters are expected to correlate non-trivially with the cosmological parameters, the tight relationships shown in Fig.\,\ref{fig:bias_fits} imply that tighter cosmological bounds could be obtained using a simulation-informed prior than the simple PT form -- this will be tested below.

\subsection{Comparing Large-Scale, Small-Scale, and Joint Constraints}
\begin{figure}[H]
    \centering
    \includegraphics[width=0.8\linewidth]{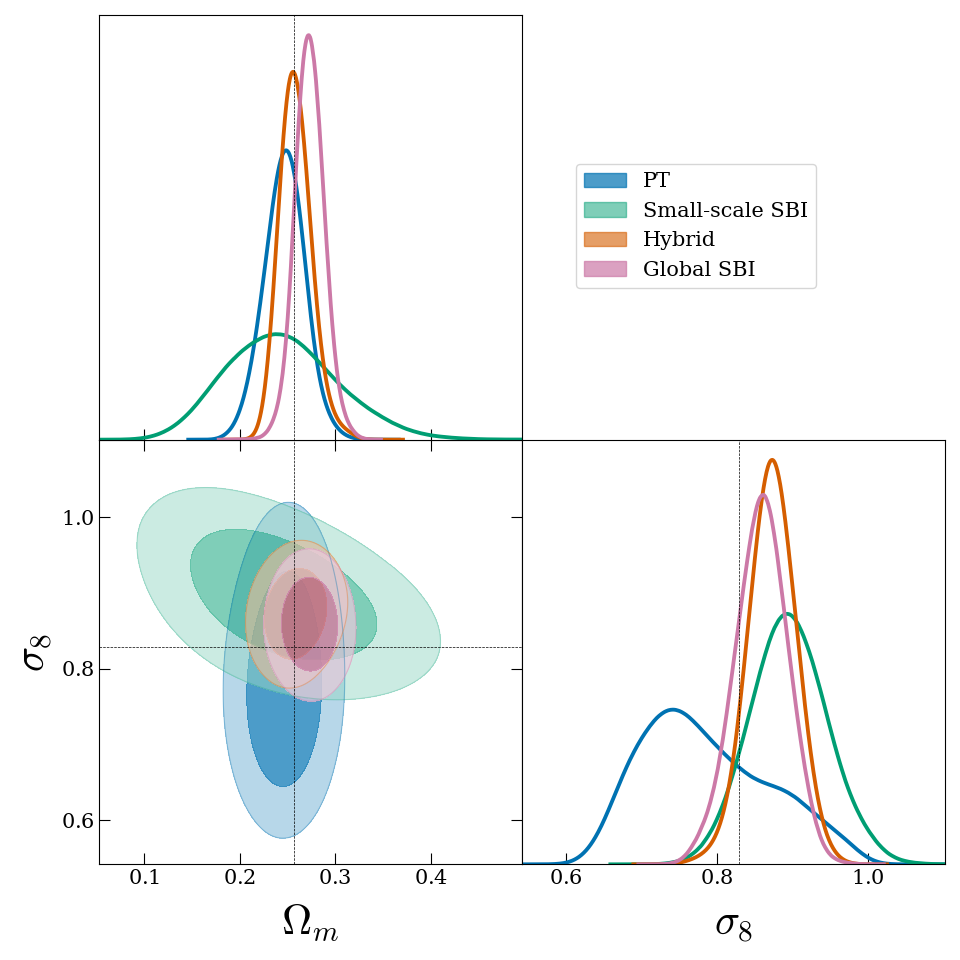}
    \caption{68th and 95th percentile contours on $\Omega_m$ and $\sigma_8$ inferred from redshift-space galaxy power spectra using large-scale PT (blue), small-scale SBI (green), HySBI (orange, new to this work), and SBI on all scales (magenta). HySBI outperforms both large-scale PT and small-scale SBI and is comparable to the global SBI result, despite being trained only on small-volume simulations. This demonstrates the importance of combining both large- and small-scale information in galaxy survey analysis.}
    \label{fig:largevssmall}
\end{figure}

\noindent In Fig.~\ref{fig:largevssmall}, we show constraints on $\Omega_m$ and $\sigma_8$ from the HySBI analysis of redshift-space galaxy power spectra.
These results use the simulation-informed prior for the bias parameters and counterterms conditioned on cosmology and HOD parameters. The impact of this prior choice will be discussed in Sec.~\ref{sec:compare_hybrid}.
We can compare these results with constraints obtained using (a) only large-scale spectra modeled with PT,  (b) only small-scale spectra modeled with SBI, and (c) both large- and small-scale spectra modeled using SBI. Although the goal of the HySBI program is to facilitate analyses in regimes where it is not feasible to run
large-volume high-fidelity simulations, 
the latter setup serves as our optimal baseline, indicating the maximum amount of information that can be extracted from the mock observations. 

From the figure, we see that significantly stronger cosmological constraints can be obtained by combining small- and large-scale statistics using HySBI. In particular, we outperform both PT on large scales ($\sim$20\% improvement on $\Omega_m$ and 60\% on $\sigma_8$) and SBI on non-linear scales ($\sim$70\% on $\Omega_m$ improvement and 40\% on $\sigma_8$). As expected, HySBI yields much stronger constraints on $\Omega_m$ than small-scale SBI while gains over PT are more moderate; this is because the large-scale shape of the linear power spectrum (including the baryon acoustic oscillations and turnover feature) contains most of the information to constrain $\Omega_m$ \citep[e.g.,][]{Philcox:2020xbv}.\footnote{We find a similar effect for $H_0$. This is omitted from the figures, however, since it is dominated by large-scale baryon acoustic oscillation information, and thus not a useful test of our hybrid pipeline.} On the other hand, changes in $\sigma_8$ manifest on all scales of the power spectrum with a stronger effect in the non-linear regime, where simulations can break the degeneracy with galaxy formation physics present in the large-scale regime (up to redshift-space effects) and the cosmic variance is smaller (as in \citealt{Hahn:2023udg}). This is shown explicitly in Appendix \ref{app}. As a result, HySBI significantly outperforms PT in constraining $\sigma_8$ and is slightly more constraining than small-scale SBI. 

Finally, we note that global SBI (trained on power spectra spanning the entire $k$ range) and HySBI lead to comparable constraints, indicating that HySBI is successful at capturing information from all scales of observation and our simulation-informed prior is able to offset the cost of constraining additional model parameters in our hybrid analysis. We discuss this in more detail in the next section. 

Taken together, the results highlight the importance of combining information from both large- and small-scales of galaxy surveys. In particular, in the absence of sufficiently large high-fidelity simulations, HySBI offers a promising avenue to more optimally extract information from future surveys -- we are able to \textit{significantly} enhance the small-scale posterior without running large-volume simulations.

\subsection{The Role of Priors in HySBI}
\label{sec:compare_hybrid}
\noindent Having demonstrated that HySBI performs comparably to global SBI, we now investigate the effect of the HySBI prior that controls the interplay of small- and large-scale information. As discussed in Sec.~\ref{sec:hybrid_sbi}, we adopt broad, independent priors on the cosmology parameters ($\theta_\rmC$) and HOD parameters ($\theta_\rmS$) in all cases. For the large scale EFT parameters ($\theta_\rmL$) which consist of bias parameters ($\theta_{\rm bias}$), and counterterms/stochastic terms ($\theta_{\rm counter}$), which marginalize over small-scale physics, including baryonic effects, we consider the following three cases:
\begin{enumerate}
    \item Conservative priors on all EFT parameters, as introduced in Sec.~\ref{sec:PT}. This neglects any correlations between large- and small-scale model parameters. 
    \item A simulation-informed prior on all EFT parameters, conditioned on the other parameters \textit{i.e.}~$p(\theta_\rmL|\theta_{\rm S},\theta_{\rm C}) = p(\theta_{\rm bias}, \theta_{\rm counter} | \theta_{\rm S}, \theta_{\rm C})$. This fully accounts for the interplay between the large- and small-scale nuisance parameters.     
    \item An intermediate scenario that models the simulation-informed prior only for the bias parameters but assumes broad independent prior on the counterterms, \textit{i.e.}~$p(\theta_\rmL) = p(\theta_{\rm bias} | \theta_{\rm S}, \theta_{\rm C}) p(\theta_{\rm counter})$. This is motivated by noting that the inference of bias parameters from simulations is much more robust than that of the counterterms (which absorb a number of additional effects, including the simulation resolution artefacts and baryonic effects).
\end{enumerate}

\begin{figure}
    \centering
    \includegraphics[width=0.8\linewidth]{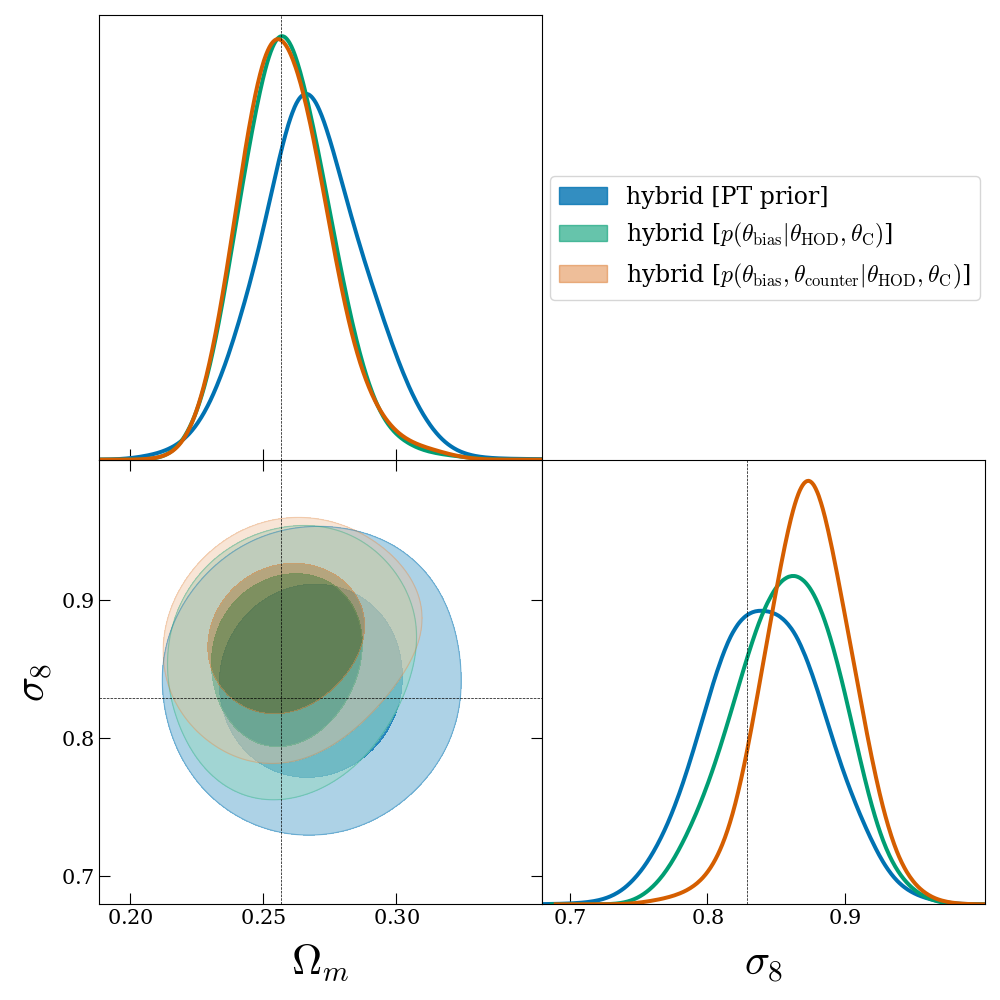}
    \caption{Comparison of HySBI posteriors using three choices of priors on large- and small-scales: a conservative PT prior that decouples the two sets of parameters (blue), a hybrid prior combining a conservative prior on counterterms and stochasticity with a simulation-informed prior on the four bias terms, conditioned on cosmology and HOD parameters (green), and a simulation-informed prior on all EFT parameters conditioned on cosmology and HOD (orange). This shows that the simulation-informed prior extracts more information than a conservative PT prior, demonstrating the importance of jointly modeling large- and small-scale parameters in HySBI.}
    \label{fig:compare_hybrid}
\end{figure}

 In Fig.~\ref{fig:compare_hybrid} we show constraints on $\Omega_m$ and $\sigma_8$ for the HySBI approach using these three priors. 
As expected, using the simulation-informed priors leads to greater constraining power on both $\Omega_m$ and $\sigma_8$ relative to the conservative PT priors. Interestingly, we observe that the constraints on $\Omega_m$ do not change if we switch from informed to conservative priors on the counterterms, as long as we use informed priors on the bias parameters. On the other hand, using informed priors for counterterms significantly improves constraints on $\sigma_8$. We expect this to happen for the same reasons as in the previous section -- the information content on $\Omega_m$ is dominated by large scales where counterterms play a small role, while $\sigma_8$ is also informed by small scales where they become important. 
However, the main conclusion from the figure is clear: joint priors on large- and small-scale parameters are essential to extracting maximal cosmological information in hybrid galaxy clustering analyses.

\subsection{Flow Priors for PT Analyses}
\label{sec:compare_prior_pt}
\noindent Having investigated the importance of simulation-informed priors for HySBI, we now briefly examine the effect of priors in the large-scale PT-only analysis. We seek to answer the following question: could we get more out of large-scale analyses by folding in simulation-informed priors? To do this, we run experiments with four different choices of the EFT parameter prior, again splitting into $\theta_{\rm bias}$ (bias) and $\theta_{\rm counter}$ (counterterms and stochastic terms): 
\begin{enumerate}
    \item Conservative priors (as defined in Sec.~\ref{sec:PT}), emulating most previous PT analyses.
    \item An intermediate case of a simulation-informed prior on the bias parameters, $p(\theta_{\rm bias} | \theta_{\rm C})$ with conservative priors on $\theta_{\rm counter}$. 
    \item A simulation-informed prior on both bias and counterterms, $p(\theta_{\rm bias}, \theta_{\rm counter} | \theta_{\rm C})$. This assumes that all EFT parameters can be robustly modeled using simulations.
    \item An extremely informative scenario of a flow prior approximating $p(\theta_{\rm bias}, \theta_{\rm counter} | \theta_{\rm C, truth})$, where $\theta_{\rm C,truth}$ is the set of true cosmological parameters for the simulation under study. This matches (3) but does not account for the cosmology-dependence of the prior, and is designed to emulate previous fixed-cosmology works \citep[e.g.,][]{ivanov2024fullshape2}.\footnote{Here, we take the simulation-informed prior trained for 3) conditioned on cosmology and fix cosmology parameters to their true values during sampling.}
\end{enumerate} 

\begin{figure}[]
    \centering
    \includegraphics[width=0.49\linewidth]{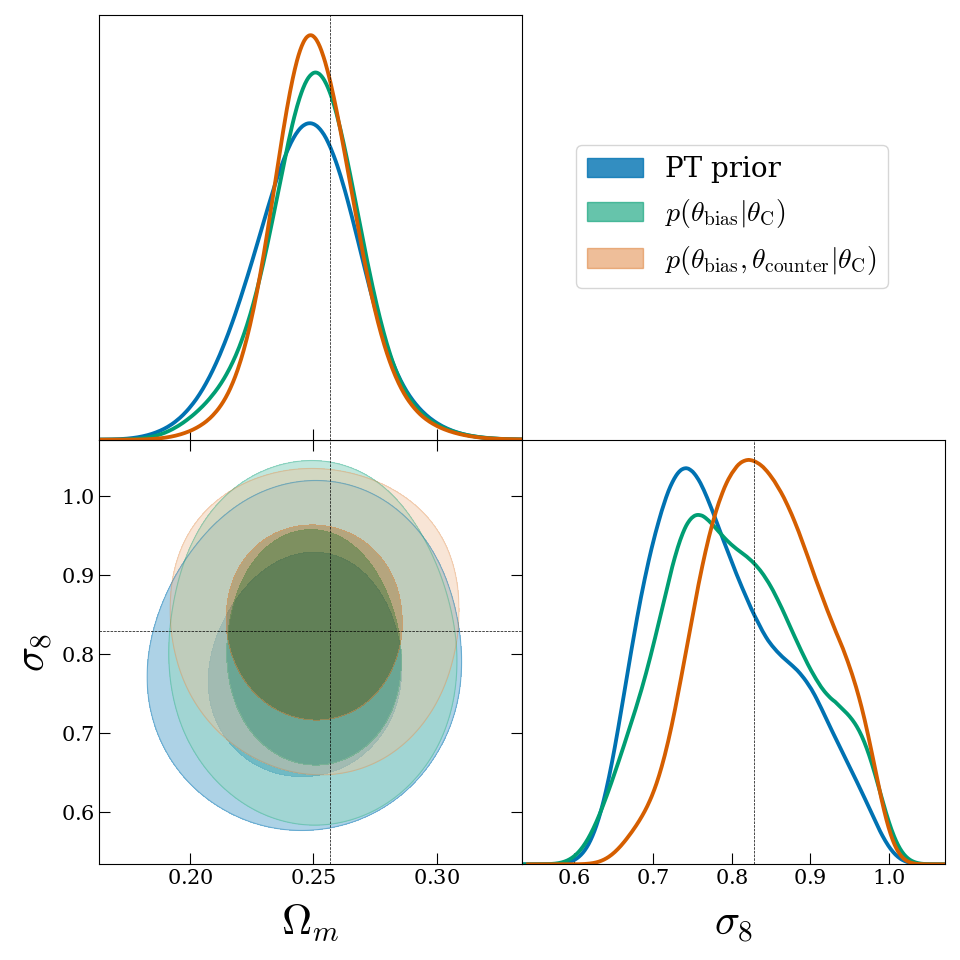}
    \includegraphics[width=0.49\linewidth]{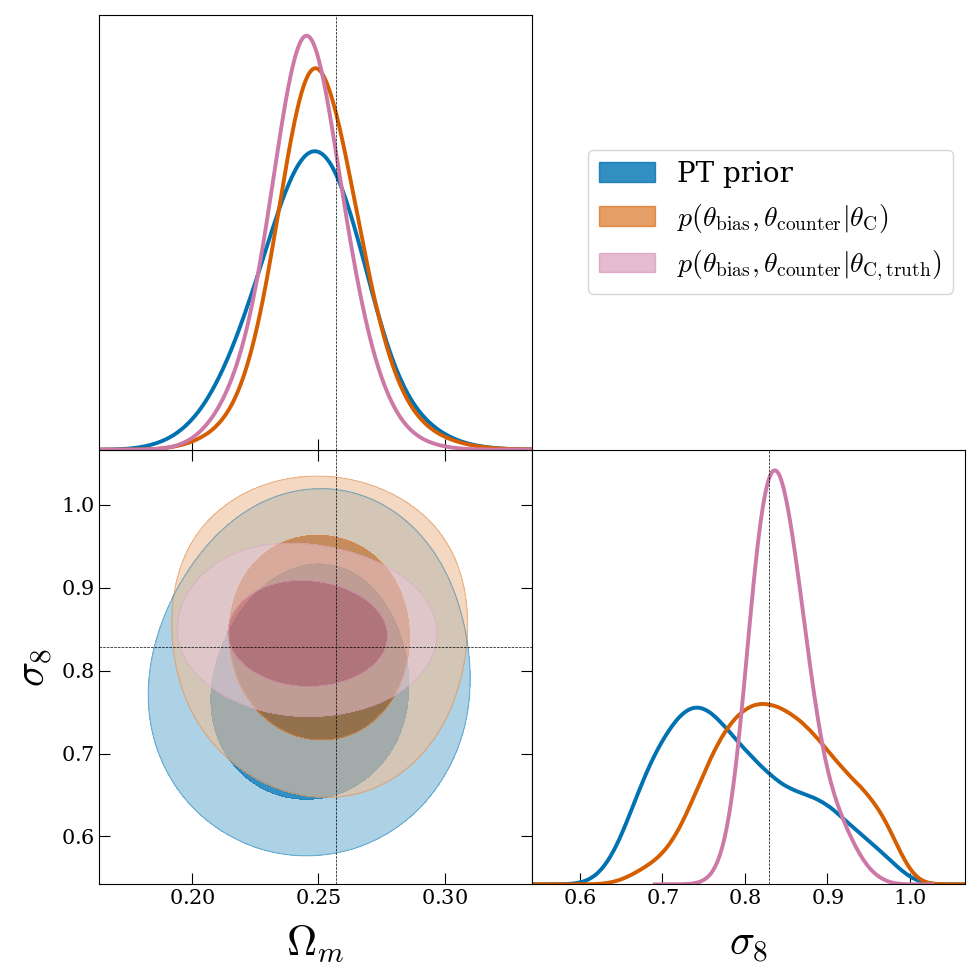}
    \caption{Comparison of the large-scale PT posteriors 
    obtained using different choices of nuisance-parameter priors. In the left panel, we compare a conservative PT prior (blue), a simulation-informed cosmology-dependent flow prior on biases with a conservative prior on all other EFT parameters (green), and a cosmology-dependent flow prior on all EFT parameters (orange). In the right plot, we compare the conservative PT prior with flow priors on all EFT parameters either conditioned on cosmology (orange) or fixing to the true cosmology (magenta). We find that the physically-motivated flow prior leads to tighter constraints on the cosmological parameters than the conservative PT prior. Furthermore, we a significant (artificial) tightening of the $\sigma_8$ constraint when conditioning the flow prior on the true cosmological parameters of the mock observation as opposed to using cosmology-dependent flow priors.}
    \label{fig:ptflow}
\end{figure}

The impact of these priors on large-scale PT analyses is shown in Fig.~\ref{fig:ptflow}. 
Notably, the simulation-informed flow prior constrains the EFT parameters to a smaller space
of physically realistic combinations of parameters, thus shrinking the effective prior volume as compared to the conservative, uninformed PT priors. 
As such, the left panel of Fig.~\ref{fig:ptflow} demonstrates that the use of informed priors leads to tighter constraints on $\Omega_m$ and $\sigma_8$ than possible with conservative priors. 
We find a slight improvement of the posteriors when including the counterterm and stochasticity parameters in the simulation-informed prior. This indicates that physics-informed priors on these parameters (which are often omitted in simulation-based treatments e.g., \citealt{Akitsu:2024lyt,Lazeyras:2017hxw}) could add some cosmological information. 
In the right-hand panel of Fig.~\ref{fig:ptflow}, we show the results using the fourth prior described above which assumes perfect knowledge of the true cosmology. Although this scenario cannot be realized when analyzing observational data, it serves as a useful test case for comparison. When conditioning our prior on the true cosmology, we observe a 26\% and 61\% fractional improvement in $\Omega_m$ and $\sigma_8$ constraints respectively, which is consistent with \cite{ivanov2024fullshape}. 

Taken together, these two results imply that training an EFT prior at fixed cosmology can result in over-confident posteriors. This is in some disagreement with the results of \cite{ivanov2024fullshape2} and \cite{zhang2024hodinformed}, which noted that there is minimal cosmology dependence in the HOD-based EFT prior. However, we caution that our analysis uses a somewhat different HOD model, 
a much wider range of cosmological parameters, and different galaxy number density cuts.\footnote{HOD parameters and cosmology parameters are degenerate in simplified settings (see \cite{ivanov2025}), but here the cosmology dependence of EFT parameters may be due to effects not captured by a simplified halo model.} We leave a systematic study of this for future work.


In summary, we find that utilizing informed priors on EFT parameters from simulations for a given galaxy sample can improve constraints on cosmology parameters even from large-scale only analysis with EFT, particularly when one imposes a prior on counter-terms as well as biases. We caution, however, that the specific gains depend on the choice of HOD model and whether one fully incorporates the cosmology dependence of these parameters. 



\vskip 20pt
\section{Discussion \& Conclusions}
\label{sec:conclusion}
\noindent SBI provides a promising avenue to complement analytical perturbative methods, allowing for full exploitation of the cosmological information present in small-scale galaxy clustering. However, with current and upcoming wide-field surveys probing an increasingly large volume of our Universe, it will be difficult to scale up the necessary high-fidelity simulations. To address this, \cite{modi2023hybridsbi} presented a hybrid framework (HySBI) that combines PT on large scales with SBI on small scales, bypassing the need for large-volume high-resolution simulations. Furthermore, they demonstrated that HySBI achieves promising results on a simple test-case: inferring cosmology parameters from real-space dark matter density fields. 

In this work, we take a crucial step towards the application of HySBI on observational data by applying it to galaxy clustering. To do this, we make the following extensions to the HySBI pipeline: (1) adding redshift-space distortions and the associated power spectrum multipoles; (2) building a pipeline for the combined modeling of galaxies on large and small scales (using EFT on large scales and HODs on small scales); (3) jointly modeling the nuisance parameters of the EFT and HOD through a simulation-learned prior. We apply HySBI to cosmological inference using the galaxy power spectra obtained from HOD realizations of the \textsc{quijote} high-resolution Latin-hypercube simulations. We also examine the impact of using a simulation-informed prior on the EFT parameters of both large-scale PT analyses and the HySBI framework.

We summarize several key findings from our work that will help inform future galaxy clustering analyses: 
\begin{enumerate}
    \item HySBI leads to tighter constraints on $\Omega_m$ and $\sigma_8$ than either small-scale SBI or large-scale PT. This makes it critical to utilize information from both scales when inferring cosmological parameters from galaxy survey data.
    \item Although the HySBI framework employs more nuisance parameters than PT- or SBI-only analyses (since it involves both large- and small-scale parameters), we can fully compensate for any loss of cosmological information by using simulations to learn an informed prior on the EFT parameters, conditioned on cosmology and small scale model parameters.
    We show that with these informed priors, HySBI has the potential to achieve comparable constraints on cosmology parameters to global SBI, but at reduced computational cost. 
    \item We show that, after marginalizing over the HOD parameters, these informative priors on EFT parameters can be used to strengthen constraints in PT analyses with respect to conservative PT priors. Differently from previous works, we find that the simulation-informed EFT priors have considerable cosmology-dependence, which one can account for by conditioning the prior during inference. 
\end{enumerate}

An important caveat about (3) is that, since the EFT priors are learned from simulations, they also become susceptible to the modeling assumptions made by these simulations. As such, the information gain during inference relies on the assumption that the simulations realistically emulate the observation. Of course, this assumption is always employed in SBI contexts; in PT applications, however, part of the attraction of conservative priors is that we are (roughly) indifferent to the precise details of the galaxy properties \citep[e.g.,][]{Modi23_sensitivity}.

While this work builds the HySBI pipeline for galaxy clustering analyses using simulated data, more work remains for application to observational data. First, HySBI requires running both small- and large-scale simulations with the same initial conditions. In this paper, we emulate this by splitting \textsc{Quijote} simulations in to subboxes; however, we are limited to analyzing observations up to the simulation box volume. To scale HySBI up for large observations, it will be important to develop methods that can produce sub-volume simulations which are consistent with the data used on large scales. Second, we note that the HySBI franework could be extended significantly beyond the HOD modeling adopted herein. In particular, one could include more nuanced simulations such as galaxy formation models that include baryonic effects. Third, HySBI can be extended to other summary statistics beyond the power spectra considered in this work, such as the wavelet scattering coefficients considered in~\cite{modi2023hybridsbi}. 
Testing HySBI with other summary statistics will be a useful step towards ensuring its robustness. 
Finally, in order to apply to observations, we must consider a number of survey systematics, such as masks, redshift distributions and fiber collisions. Incorporating these effects in simulations will be important for the application of HySBI on future surveys. 

\begin{acknowledgments}
\noindent We thank Mikhail Ivanov for insightful discussions. GZ was supported by the National Science Foundation under Cooperative Agreement PHY-2019786 (The NSF AI Institute for Artificial Intelligence and Fundamental Interactions, http://iaifi.org/). CM is supported by the James Arthur Fellowship at NYU. OHEP is a Junior Fellow of the Simons Society of Fellows. The computations in this work were run on the FASRC Cannon cluster supported by the FAS Division of Science Research Computing Group at Harvard University and at facilities supported by the Scientific Computing Core at the Flatiron Institute, a division of the Simons Foundation.

\end{acknowledgments}

\pagebreak
\appendix 
\section{Comparing HySBI with Flow-Prior PT}\label{app}
\begin{figure}[H]
    \centering
    \includegraphics[width=0.7\linewidth]{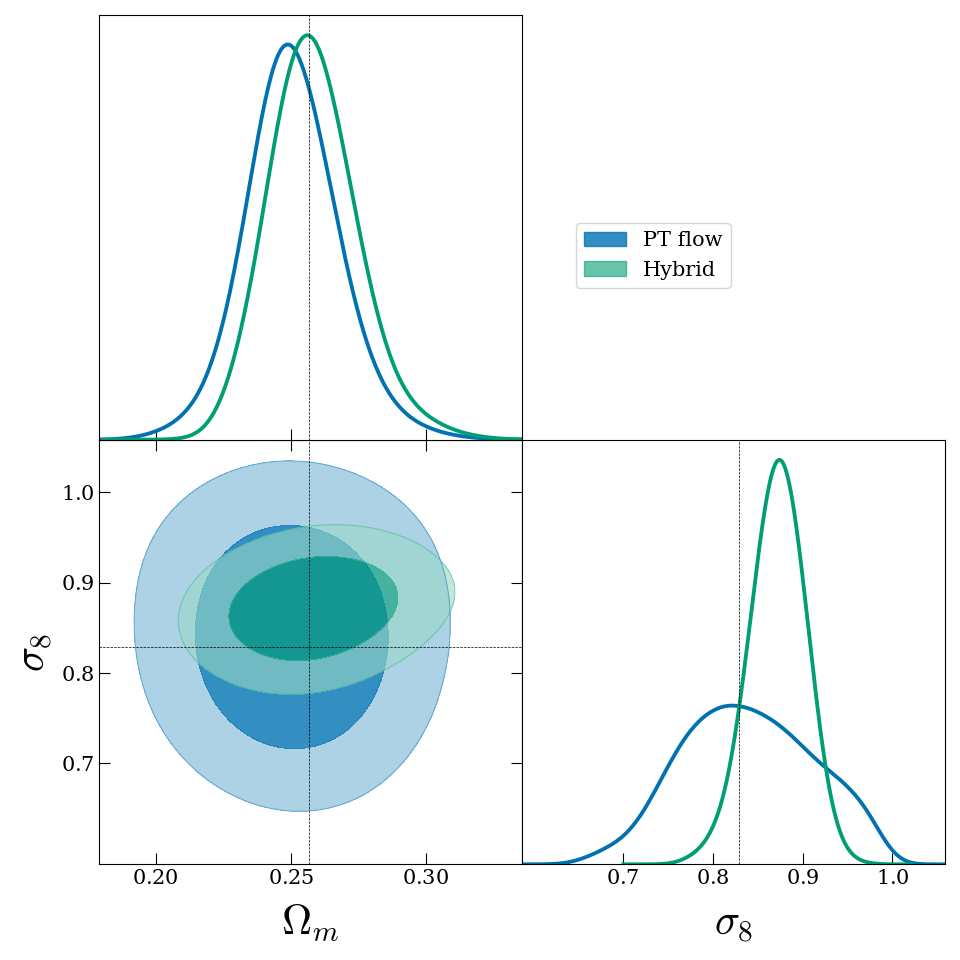}
    \caption{Comparison of posterior distributions obtained from two analyses using a flow prior on bias parameters and counterterms: large-scale PT analyses (blue) and combined HySBI analyses (green). While the constraints on $\Omega_m$ are dominated by large scales, we find significant tightening in the $\sigma_8$ posterior when small scales are included using our HySBI method.}
    \label{fig:ptflowhybrid}
\end{figure}

\noindent To determine which scales of the galaxy power spectrum are most useful for probing $\sigma_8$, we compare the constraints from two analyses in Fig.~\ref{fig:ptflowhybrid}: (1) PT using a SBI prior (as shown in Fig.~\ref{fig:ptflow}); (2) HySBI (as shown in Fig.~\ref{fig:compare_hybrid}) . We find that the large-scale PT analysis provides much weaker $\sigma_8$ constraints than the hybrid pipeline, despite the strong improvements sourced by the simulation-informed prior. This implies that small-scale statistics dominate the information content of $\sigma_8$ in HySBI, even with a shot-noise-limited galaxy sample. 
\label{app:ptflow_vs_hybrid}

\bibliography{references}

\end{document}